**Comet 21P/Giacobini-Zinner: Water Production Activity over 20 Years with SOHO/SWAN**

Short Title: SOHO/SWAN Observations of comet 21P

M.R. Combi[1], T. Mäkinen[2], J.-L. Bertaux[3], E. Quémerais[3], S. Ferron[4] and R. Coronel[1]

[1]Dept. of Climate and Space Sciences and Engineering
University of Michigan
2455 Hayward Street
Ann Arbor, MI 48109-2143
*Corresponding author: mcombi@umich.edu

[2]Finnish Meteorological Institute, Box 503
SF-00101 Helsinki, FINLAND

LATMOS/IPSL
Université de Versailles Saint-Quentin
11, Boulevard d'Alembert, 78280, Guyancourt, FRANCE

ACRI-st, Sophia-Antipolis, FRANCE

**ABSTRACT**

In 1985 Comet 21P/Giacobini-Zinner was the first comet visited by a spacecraft, the International Cometary Explorer (ICE) satellite, several months before the armada of Halley spacecraft had their encounters in 1986. ICE was originally the ISEE-3 satellite, designed for magnetospheric measurements near the Earth, and was diverted via a lunar gravity assist to pass through the plasma tail of the comet. The comet has been observed by the all-sky hydrogen Lyman-alpha Solar Wind Anisotropies (SWAN) camera on the SOlar and Heliospheric Observatory satellite during its last four apparitions in 1998, 2005, 2012 and 2018. This paper compares water production rates calculated from the hydrogen images from the 1998 and 2005 results, published by Combi et al. (2011), with new observations from 2012 and 2018. Unlike some comets that have faded over time, except for 2 outbursts seen in the 2012 results, the activity levels for Comet 21P have not changed consistently over time. A power-law fit to the pre-perihelion water production rate vs. heliocentric distance for all four apparitions gives a result of $3.75 \times 10^{28} \times r^{-0.28 \pm 0.4}$, and that for the post-perihelion production rate, discounting 2012 data dominated by outbursts, gives $4.15 \times 10^{28} \times r^{-10.2 \pm 0.9}$. The production rates are in $s^{-1}$ and heliocentric distances are in AU.

## 1. Introduction

Comet 21P/Giacobini-Zinner (hereafter G-Z) was discovered in 1900 by Michel Giacobini and recovered in later apparitions by Ernst Zinner. It has been widely observed on nearly every apparition ever since. The plasma tail of G-Z was the first comet visited by a spacecraft, the International Cometary Explorer (Cowley 1985), in 1985, which had a complement of particle and field instruments, including instrumentation to make ion composition measurements (Ogilvie et al. 1986). The photometric survey of comets by A'Hearn et al. (1995) classified G-Z as the archetype member of a group of comets that is depleted in $C_2$ and $C_3$ and slightly depleted in NH, all relative to CN. Fink (2009) found it similarly depleted in $C_2$ and $C_3$ and also depleted in $NH_2$. The agreement of the small depletion of NH and $NH_2$ implies a depletion of the parent molecule $NH_3$.

G-Z has also been observed in the UV and IR. During the 1985 apparition, G-Z was observed with the International Ultraviolet Explorer satellite, and water production rates were determined from observations of OH at 309 nm (McFadden et al. 1987) and H Lyman-alpha (Combi and Feldman 1992). H Lyman-alpha was also observed with the Pioneer Venus UVS instrument (Combi et al. 1986). These results were summarized and compared with the 1998 and 2005 apparitions of H Lyman-alpha observed with the Solar Wind Anisotropies instrument (SWAN) on the SOlar and Heliospheric Observatory (SOHO) satellite by Combi et al. (2011).

Infrared observations by Weaver et al. (1999) and Mumma et al. (2000) showed that, like the carbon-chain radical depletion compared to OH in the visible, $CH_3OH$ and $C_2H_6$ were depleted compared to water. Later observations by DiSanti et al. (2013) confirmed these depletions. Mumma et al. (2000) found a rather substantial $CO/H_2O$ ratio in the range of 6%-15%, in contrast to a much lower upper limit by Weaver et al. (1999). All measurements summarized by Dello Russo et al. (2016) give a weighted average value for $CO/H_2O$ of only (2.2 ± 1.5)%. Shinnaka et al. (2020) found a very low $CO_2/H_2O$ ratio in G-Z of only 1-2% and only upper limits or non-detections were available for other species, such as HCN, $NH_3$, $C_2H_2$ and $CH_4$. However, a recent and more sensitive set of IR observations of G-Z give improved measurements (Roth et al. 2020) of CO (between 1% and 2%), $CH_4$ (between 0.6% and 1.1%) and $C_2H_6$ (between 0.24% and 0.29%) compared to water. They also found some variation over the orbit of G-Z, indicating some possible compositional heterogeneity in the innermost ±1000 km of the coma toward either side of the nucleus, indicating possible seasonal effects comparing pre-, near-, and post-perihelion observations.

## 2. SOHO/SWAN Observations and Basic Analysis

The SWAN instrument on board the SOHO satellite makes nearly daily whole sky observations of the hydrogen Lyman-$\alpha$ 121.5-nm emission. As such, its major investigation is to measure the 3-dimensional structure of the solar wind and solar radiation fields via their interaction and resulting depletion of the interstellar neutral hydrogen streaming through the solar system (Bertaux et al. 1995). The sensitivity of SWAN, as well as its location on the SOHO spacecraft in a halo orbit around the Sun-Earth L1 Lagrange point, make it an ideal observatory to view the large atomic hydrogen comae surrounding comets (Bertaux et al. 1998; Combi et al. 2019). SWAN enables daily 1° by 1° resolution imaging of bright enough comets (typically brighter than $m_v \sim 10$) independent of north/south hemisphere location or (mostly) the relative positions of the comet, Sun and Earth, a definite advantage over observations made from the ground or from low Earth orbit. Other impediments to obtaining water production rates from comet observations or affecting the resulting uncertainties are caused by one or more of the following: bright stars near the comet (especially when the comet in is or near the galactic equator), SWAN's solar avoidance area, or when portions of the spacecraft block the line-of-sight to the comet.

Because photodissociation of water and its main daughter radical, OH, produces two H atoms per water molecule, modeling an observation of the H coma and its variation along a comet's orbit has proven to yield a consistent set of water production rates for many comets. The methodology, our time-resolved model (or TRM), as well as background theoretical and observationally-verified parameterizations have been explained in detail and documented in several previous papers (Combi et al. 2004; Mäkinen and Combi 2005; Combi et al. 2005; Combi et al. 2019) and need not be discussed in detail yet again here.

Total uncertainties in SWAN water production rates result from the fitting process of the comet model and the background interplanetary hydrogen to the observation, noise/uncertainty in the measurement as well as brightness calibration and errors associated with the model and model parameters. The smaller listed and plotted 1-σ uncertainties result only from the noise in the data and the fitting process. Total uncertainties resulting from the SWAN calibration, the solar measurements of integrated Lyman-alpha and model associated errors are expected to be on the order of ~30%. In this paper, owing to the fact that observations are of a moderately weak Jupiter Family comet (JFC), G-Z, we give single-image water production rates obtained from each image. The filling time of the field of view used (a circular radius of 8 degrees) by hydrogen atoms streaming outward at speeds of 8-20 km s$^{-1}$, combined with the photodissociation times of $H_2O$ and OH, ~80,000 and ~150,000 s at a heliocentric distance of 1 AU and spacecraft-comet distances of ~0.4-1 AU result in the H coma representing the water production over the previous 2-3 days.

The solar Lyman-alpha g-factors are calculated from the daily disk-integrated solar Lyman-alpha fluxes from the LASP web site at the University of Colorado (http://lasp.colorado.edu/lisird/lya/), in combination with the solar Lyman-alpha velocity dependent line profile (Lemaire et al. 1998). Daily-fluxes are measured from the Earth-facing disk of the Sun, so the value at the comet, which is in a different direction, are taken from the nearest time the same face of the Sun was seen by the comet, accounting for the 27-day rotation of the Sun. Typically, variations of solar Lyman-alpha caused by active regions on the Sun repeat on subsequent rotations.

### 3. Water Production Rates in Comet 21P/Giacobini-Zinner

G-Z was observed by SWAN during the last four apparitions, 1998, 2005, 2012, and 2018. A summary of the data set is given in Table 1. The resulting water production rates, observational circumstances and ancillary data for the 1998 and 2005 apparitions were published in a paper by Combi et al. (2011) and also appear in the SOHO/SWAN database of over 70 comets in the NASA Planetary Data System Small Bodies Node (Combi 2017, 2020). The daily-average water production rates, formal model-fitting uncertainties, g-factors and observational circumstances for the 2012 and 2018 apparitions are given in Table 2.

Figure 1 shows a plot of the water production rate as a function of time in days from perihelion for all four apparitions. There is clearly some irregularity in activity within each apparition, as well as from apparition to apparition. There is, however, no evidence of consistent fading over 20 years as there has been in comets 103P/Hartley 2, 41P/Tuttle-Giacobini-Kresak and 46P/Wirtanen (Combi et al. 2020). The irregularity in activity over time is also apparent in visual magnitude variations (Yoshida 2020). In particular, the water production rate variations are noteworthy during the 2012 apparition, which has some of the larger departures from the mean activity, both during the pre- and post-perihelion legs of the orbit. In particular, there were large increases in production rate in 2012 around 34 days before perihelion and 30 days after perihelion, indicating likely outbursts at the level of a factor of two.

Table 1. Summary of SOHO/SWAN Observations, Comet 21P/Giacobini-Zinner

| Perihelion Date | q(AU) | # of Images | $r_H$ range (AU) |
| --- | --- | --- | --- |
| 21.320 Nov. 1998 | 1.024 | 18 | 1.034 – 1.108 |
| 2.761 Jul. 2005 | 1.038 | 39 | 1.038 – 1.284 |
| 11.735 Feb. 2012 | 1.030 | 19 | 1.058 – 1.213 |
| 10.278 Sep. 2018 | 1.013 | 71 | 1.013 – 1.284 |

$r_H$ = heliocentric distance (AU)
q = perihelion distance (AU)

The water production rate levels and characteristic variation peaking at about 40 days before perihelion and then decreasing through perihelion and up to 40 days after is very similar to and agrees with that reported by Schleicher et al. (1987) from photometric data of OH obtained during the 1985 apparition. Their data cover a larger range of times, spanning from approximately 80 days before to 80 days after perihelion. Before 40 days before perihelion their results show that the production rate does continue to decrease, as one might expect, but from 40 to 80 days after perihelion the OH production rate levels off at ~$10^{28}$ s$^{-1}$.

Figure 2 shows the production rates from all four apparitions plotted as a function of heliocentric distance. The previous paper that included the results from the 1998 and 2005 apparitions (Combi et al. 2011) found pre- and post-perihelion power-law fits to the production rate of 5.9 x $10^{28}$ $r^{-1.7\pm0.4}$ and 6.8 x $10^{28}$ $r^{-11.9\pm3.5}$, respectively, with r being the heliocentric distance in AU, and the production rate at 1 AU in s$^{-1}$. The combination of all four apparitions, leaving out the 2012 post-perihelion data dominated by the outburst, gives a pre-perihelion result of 3.75 x $10^{28}$ x $r^{-0.28\pm0.4}$, and for the post-perihelion production rate give 4.2 x $10^{28}$ x $r^{-10.2\pm0.9}$. It appears that the large outburst values from 2012 significantly influence the power laws. While the pre-to-post perihelion asymmetry remains similar, with a much flatter slope before perihelion and a very steep slope after. The post-perihelion fit for just the 2018 is 4.0 x $10^{28}$ $r^{-9.2\pm1.4}$, which is to within uncertainties the same as that for 1998, 2005 and 2018.

It is notable that there is a decided change in slope (exponent) separating the pre-perihelion data before -40 days. After -40 days the production rate is essentially flat, while before -40 days the production rates decrease as one might expect, and as is seen in the ground-based photometry of Schleicher et al. (1987). If we fit a power law only to the data before -40 days we obtain a slope of -3.9, or an expression of $7.07 \times 10^{28} r^{-3.9 \pm 1.0}$. This is a more typical drop in production rate with heliocentric distance compared with other Jupiter Family Comets (Combi et al. 2019) and highlights the strong seasonal variation of water production rate.

Table 2

SOHO/SWAN Observations of Comet 21P/Giacobini-Zinner and Water Production Rates

| ΔT (Days) | r (AU) | Δ (AU) | g ($s^{-1}$) | Q ($10^{28}$ $s^{-1}$) | δQ ($10^{28}$ $s^{-1}$) |
|---|---|---|---|---|---|
| 2012 | | | | | |
| -45.890 | 1.213 | 1.974 | 0.001864 | 4.16 | 0.48 |
| -43.890 | 1.199 | 1.964 | 0.001889 | 5.23 | 0.39 |
| -42.890 | 1.192 | 1.959 | 0.001884 | 3.11 | 0.61 |
| -41.890 | 1.186 | 1.954 | 0.001850 | 5.35 | 0.37 |
| -38.890 | 1.166 | 1.940 | 0.001768 | 4.94 | 0.45 |
| -36.890 | 1.153 | 1.931 | 0.001759 | 4.19 | 0.53 |
| -34.886 | 1.141 | 1.922 | 0.001774 | 5.84 | 0.54 |
| -33.886 | 1.136 | 1.918 | 0.001789 | 7.38 | 0.47 |
| -32.886 | 1.130 | 1.914 | 0.001912 | 6.10 | 0.53 |
| -31.886 | 1.124 | 1.910 | 0.001852 | 5.27 | 0.46 |
| -29.886 | 1.113 | 1.903 | 0.001824 | 5.64 | 0.31 |
| -27.916 | 1.103 | 1.896 | 0.001899 | 4.67 | 0.45 |
| -22.824 | 1.080 | 1.880 | 0.001886 | 4.50 | 0.44 |
| -21.824 | 1.076 | 1.877 | 0.001877 | 4.91 | 0.25 |
| -16.824 | 1.058 | 1.865 | 0.001866 | 6.01 | 0.27 |
| 29.158 | 1.110 | 1.919 | 0.001806 | 4.30 | 0.31 |
| 31.158 | 1.120 | 1.928 | 0.001809 | 3.14 | 0.51 |
| 37.159 | 1.155 | 1.958 | 0.001811 | 1.48 | 1.12 |
| 38.159 | 1.161 | 1.964 | 0.001810 | 1.56 | 1.06 |
| 2018 | | | | | |
| -56.567 | 1.284 | 0.713 | 0.001492 | 2.04 | 0.26 |
| -55.567 | 1.276 | 0.705 | 0.001499 | 2.45 | 0.20 |
| -54.567 | 1.268 | 0.697 | 0.001521 | 2.49 | 0.20 |
| -53.567 | 1.260 | 0.688 | 0.001513 | 2.08 | 0.22 |
| −52.567 | 1.252 | 0.680 | 0.001509 | 2.29 | 0.19 |
| −51.567 | 1.244 | 0.672 | 0.001515 | 2.82 | 0.17 |
| −50.568 | 1.237 | 0.664 | 0.001509 | 3.15 | 0.14 |
| −49.554 | 1.229 | 0.656 | 0.001506 | 2.02 | 0.27 |
| −48.554 | 1.221 | 0.648 | 0.001517 | 3.26 | 0.16 |
| −47.554 | 1.214 | 0.640 | 0.001514 | 4.54 | 0.10 |
| −46.554 | 1.206 | 0.632 | 0.001513 | 4.41 | 0.21 |
| −45.554 | 1.199 | 0.624 | 0.001499 | 4.83 | 0.11 |
| −44.554 | 1.192 | 0.616 | 0.001486 | 5.19 | 0.07 |
| −43.554 | 1.185 | 0.608 | 0.001468 | 4.77 | 0.23 |
| −42.554 | 1.178 | 0.600 | 0.001458 | 4.17 | 0.14 |
| −41.539 | 1.171 | 0.592 | 0.001450 | 3.80 | 0.13 |
| −40.539 | 1.164 | 0.584 | 0.001456 | 3.34 | 0.14 |
| −39.539 | 1.157 | 0.576 | 0.001462 | 3.92 | 0.13 |
| −38.539 | 1.151 | 0.569 | 0.001446 | 4.35 | 0.12 |
| −37.539 | 1.144 | 0.561 | 0.001451 | 4.01 | 0.14 |
| −36.525 | 1.138 | 0.554 | 0.001446 | 4.45 | 0.11 |
| −35.525 | 1.132 | 0.546 | 0.001451 | 4.11 | 0.28 |
| −34.525 | 1.125 | 0.539 | 0.001453 | 3.71 | 0.27 |
| −27.160 | 1.085 | 0.486 | 0.001455 | 3.26 | 0.42 |
| −26.160 | 1.080 | 0.480 | 0.001438 | 3.71 | 0.32 |
| −25.160 | 1.075 | 0.473 | 0.001444 | 3.51 | 0.32 |
| −24.175 | 1.070 | 0.467 | 0.001434 | 2.87 | 0.11 |
| −23.175 | 1.066 | 0.460 | 0.001429 | 2.92 | 0.11 |
| −22.189 | 1.062 | 0.454 | 0.001427 | 2.93 | 0.29 |
| −21.189 | 1.057 | 0.448 | 0.001413 | 2.77 | 0.23 |
| −19.204 | 1.050 | 0.437 | 0.001402 | 3.58 | 0.18 |
| −18.217 | 1.046 | 0.432 | 0.001408 | 3.30 | 0.27 |
| −17.217 | 1.043 | 0.426 | 0.001402 | 3.64 | 0.26 |
| −16.217 | 1.039 | 0.421 | 0.001400 | 3.25 | 0.31 |
| −15.233 | 1.036 | 0.417 | 0.001421 | 2.78 | 0.24 |
| −14.246 | 1.033 | 0.412 | 0.001408 | 2.75 | 0.19 |
| −13.246 | 1.031 | 0.408 | 0.001390 | 3.19 | 0.44 |
| −12.262 | 1.028 | 0.404 | 0.001385 | 2.84 | 0.33 |
| −11.262 | 1.026 | 0.400 | 0.001379 | 2.92 | 0.20 |
| −10.275 | 1.024 | 0.396 | 0.001381 | 2.80 | 0.19 |
| −9.290 | 1.022 | 0.393 | 0.001380 | 2.73 | 0.09 |
| −8.290 | 1.020 | 0.390 | 0.001379 | 3.31 | 0.36 |
| −7.303 | 1.018 | 0.388 | 0.001381 | 3.05 | 0.38 |
| −6.319 | 1.017 | 0.385 | 0.001407 | 3.15 | 0.11 |
| −5.319 | 1.016 | 0.383 | 0.001409 | 3.03 | 0.11 |
| −4.332 | 1.015 | 0.382 | 0.001396 | 3.03 | 0.10 |
| −3.348 | 1.014 | 0.380 | 0.001388 | 2.77 | 0.23 |
| −2.361 | 1.013 | 0.379 | 0.001383 | 3.11 | 0.26 |
| −0.377 | 1.013 | 0.378 | 0.001393 | 3.48 | 0.33 |
| 0.623 | 1.013 | 0.378 | 0.001388 | 3.98 | 0.19 |
| 1.613 | 1.013 | 0.378 | 0.001396 | 3.57 | 0.23 |
| 2.603 | 1.014 | 0.379 | 0.001395 | 3.44 | 0.13 |
| 3.593 | 1.014 | 0.380 | 0.001390 | 2.88 | 0.19 |
| 4.583 | 1.015 | 0.381 | 0.001394 | 2.46 | 0.11 |
| 5.583 | 1.016 | 0.383 | 0.001390 | 3.90 | 0.36 |
| 6.583 | 1.017 | 0.385 | 0.001393 | 2.98 | 0.33 |
| 8.583 | 1.020 | 0.390 | 0.001377 | 3.31 | 0.29 |
| 10.632 | 1.024 | 0.397 | 0.001381 | 3.21 | 0.26 |
| 11.641 | 1.027 | 0.400 | 0.001373 | 2.99 | 0.08 |
| 12.657 | 1.029 | 0.404 | 0.001369 | 2.65 | 0.11 |
| 13.657 | 1.032 | 0.408 | 0.001373 | 3.70 | 0.17 |
| 14.670 | 1.035 | 0.412 | 0.001362 | 3.38 | 0.11 |
| 15.670 | 1.038 | 0.417 | 0.001366 | 3.49 | 0.11 |
| 16.685 | 1.041 | 0.422 | 0.001378 | 3.19 | 0.12 |
| 17.685 | 1.044 | 0.427 | 0.001374 | 2.29 | 0.17 |
| 19.714 | 1.052 | 0.438 | 0.001390 | 2.33 | 0.16 |
| 21.728 | 1.060 | 0.449 | 0.001384 | 1.92 | 0.16 |
| 23.384 | 1.067 | 0.459 | 0.001386 | 2.87 | 0.24 |
| 24.384 | 1.071 | 0.465 | 0.001379 | 2.81 | 0.15 |
| 28.413 | 1.091 | 0.490 | 0.001390 | 1.73 | 0.22 |
| 34.649 | 1.126 | 0.533 | 0.001398 | 1.06 | 0.43 |

Notes. DT (Days from Perihelion February 11.735, 2012; September 10.278, 2018)

r : Heliocentric distance (AU)

Δ: Comet-Spacecraft distance (AU)

g: Solar Lyman-α g-factor (photons s$^{-1}$) at 1 AU

Q: Water production rates for each image (s$^{-1}$)

δQ: internal 1-sigma uncertainties

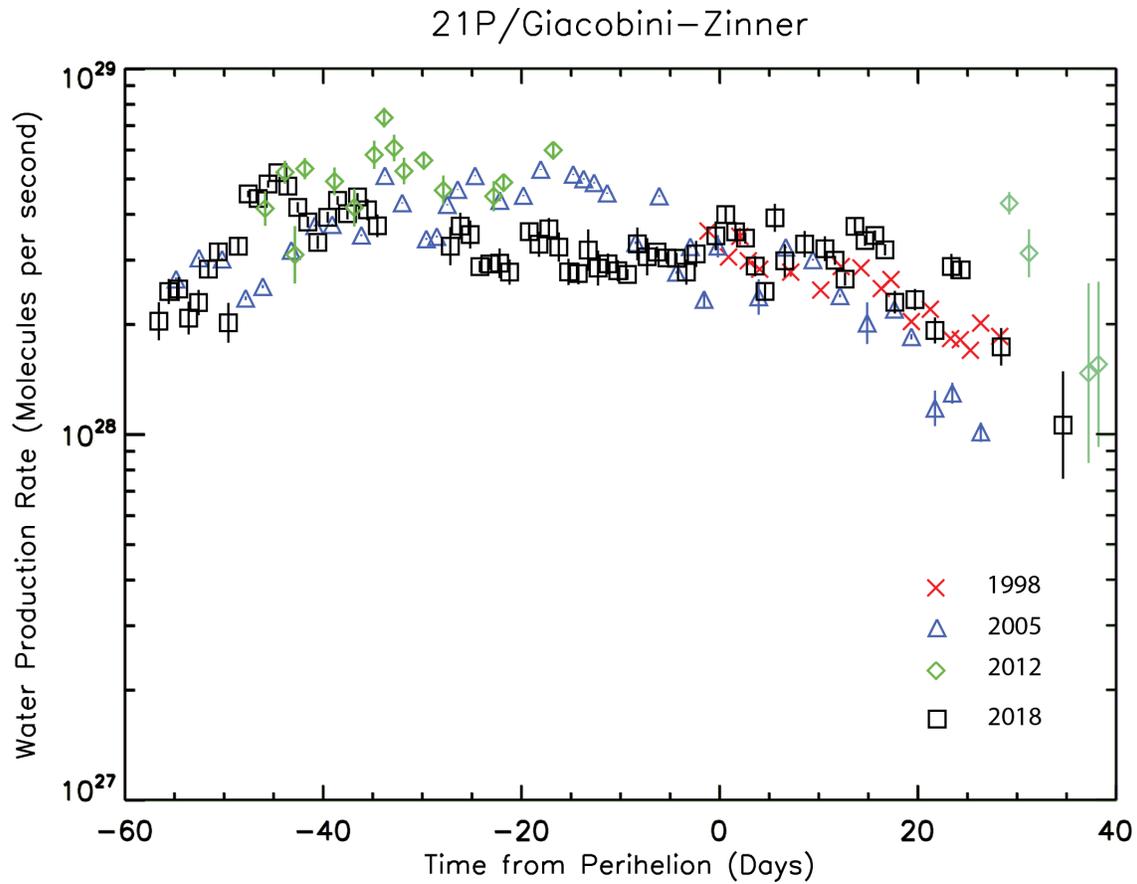

Figure 1. Single-image water production rates in comet 21P/Giacobini-Zinner are plotted as a function of time from perihelion in days. The error bars correspond to the respective 1σ formal uncertainties. The red x's show the values from the 1998 apparition, the blue triangles from 2005, the green diamonds from 2012, and the black squares from 2018. The vertical lines show the formal model fitting uncertainties.

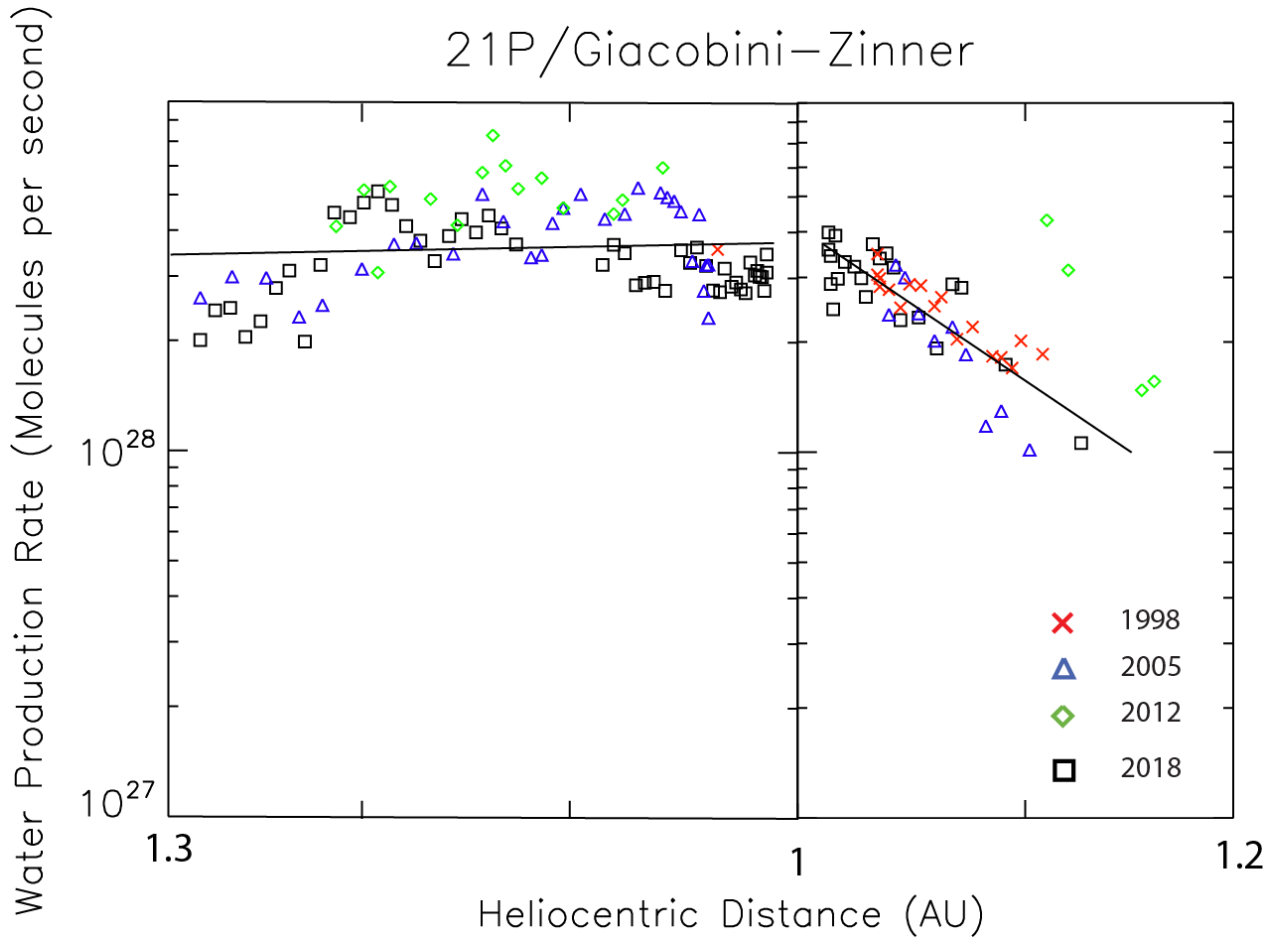

Figure 2. Single-image water production rates in comet 21P/Giacobini-Zinner are plotted as a function of heliocentric distance in AU. The pre-perihelion data are in the left half and the post-perihelion data in the right. Note that the perihelion distance is 1.03 AU and is responsible for the gap. In addition, there are small changes in perihelion distance over the 20 years. The straight lines represent the power-law fits to all four apparitions, leaving out the post-perihelion 2012 data dominated by the outburst. The best-fit coefficients for the power-law fits are given in the text. The x's (red) show the values from the 1998 apparition, the triangles (blue) from 2005, the diamonds (green) from 2012, and the squares (black) from 2018.

## 6. Summary

We present here the analysis of hydrogen Lyman-alpha observations of comet 21P/Giacobini-Zinner obtained with the all-sky SWAN camera on the SOHO spacecraft obtained during the four apparitions since the SOHO launch, 1998, 2005, 2012, and 2018. With the exception of what appear to be two outbursts during the 2012 apparition raising the water production rate by a factor of two for a short period, the water production activity appears to be quite stable over this period of time. Comparison with OH results even from the 1985 apparition look very similar.

**Acknowledgements:** SOHO is an international mission between ESA and NASA. M. Combi acknowledges support from NASA grant 80NSSC18K1005 from the Solar System Observations Program. T.T. Mäkinen was supported by the Finnish Meteorological Institute (FMI). J.-L. Bertaux and E. Quémerais acknowledge support from CNRS and CNES. We obtained orbital elements from the JPL Horizons web site (http://ssd.jpl.nasa.gov/horizons.cgi). The composite solar Lyα data was taken from the LASP web site at the University of Colorado (http://lasp.colorado.edu/lisird/lya/). We acknowledge the personnel that have been keeping SOHO and SWAN operational for over 20 years, in particular Dr. Walter Schmidt at FMI. We also acknowledge the support of R. Coronel by the Undergraduate Research Opportunity Program of the University of Michigan. We also thank the two reviewers for their careful reading and helpful suggestions that have improved the paper.